\documentclass[12pt,a4paper]{article}
\usepackage[latin1]{inputenc}
\usepackage[T1]{fontenc}
\usepackage[mathscr]{eucal}  
\usepackage{times}           
\usepackage{enumerate}
\usepackage[dvips]{graphicx}
\usepackage{fancyheadings}
\usepackage{epsfig}
\usepackage{verbatim}
\usepackage{amsmath}

\newcommand{\bd}{\begin{displaymath}}
\newcommand{\ed}{\end{displaymath}}
\newcommand{\beq}{\begin{equation}}
\newcommand{\eeq}{\end{equation}}
\newcommand{\ba}{\begin{array}}
\newcommand{\ea}{\end{array}}
\newcommand{\bal}{\begin{align}}
\newcommand{\eal}{\end{align}}
\newcommand{\bpm}{\begin{pmatrix}}
\newcommand{\epm}{\end{pmatrix}}

\begin{document}

\title{The climate version of the Eta regional forecast model. 1. Evaluation 
of consistency between the Eta model and HadAM3P global model.}

\author{I. A. Pisnichenko\footnote{Centro de Previs\~{a}o de Tempo e Estudos 
Clim\'{a}ticos/Instituto Nacional de Pesquisas Espaciais, Cachoeira Paulista, 
SP, Brazil, (pisnitch@cptec.inpe.br).  Additional affiliation: A.M. Obukhov Institute 
of Atmospheric Physics, Russian Academy of Sciences, Moscow, Russia.} 
\and T.A. Tarasova\footnotemark[1] }

\date{ 25 November 2007}
\maketitle
\centerline{\bf Abstract}

     The regional climate model prepared from Eta WS (workstation) forecast
model has been integrated over South America with the horizontal resolution 
of 40 km for the period of 1960-1990. The model was forced at its lateral
boundaries by the outputs of HadAM3P. The data of HadAM3P  represent
simulation of modern climate with the resolution about 150 km.
In order to prepare climate  regional model from the Eta forecast model
multiple modifications and corrections were made in the original model as well as 
new program blocks were added. The run of climate Eta model was made on 
the supercomputer SX-6. The detailed analysis of the results of  dynamical downscaling 
experiment includes an investigation of a consistency between the regional and AGCM 
models as well as of ability of the regional model to resolve important features 
of climate fields on the finer scale than that resolved by AGCM. In this work 
the results of the investigation of a consistency between the output fields of 
the Eta model and HadAM3P are presented. The geopotential, temperature and wind 
fields of both models were analysed. For the evaluation of the likeness of these 
two models outputs, there were used Fourier analysis of time series, consistency index, 
constituted from linear regression coefficients, time mean and space mean 
models' arithmetic difference and root mean square difference, dispersion analysis, 
and some others characteristics. This investigation demonstrates that there are not
significant differences in behaviour and spatial arragement of large-scale structures 
of the two models.  Also, the regional model characteristics do not have considerable 
positive or negative trend during integration time in relation to the global model 
characteristics. From the total analysis we can affirm that in the description of
large-scale climate structures these two models are in consistency.




\noindent {\bf 1. Introduction}  


The time averaged large-scale  meteorological  fields ( >500 km) are
actively  studied in the works on climate theory and  climate change
analysis. Needs of agriculture, industrial and energy development
planning require the knowledge of detailed, regional and local scale
(100km - 10 km) climatic information. As the modern net of climate
observation stations can supply data only suited for  large-scale
climate field investigations, the dynamical downscaling using
high-resolution regional climate model (RCM) is the most powerful
instrument for obtaining the smaller-scaled climate information. For the
study of  regional climate change in the future  the dynamical
downscaling is the only way  to obtain necessary information. The
dynamical downscaling approach involves RCM forced at the lateral and
bottom boundaries  by an atmospheric general circulation model (AGCM) or
reanalysis data (e.g., Dickinson et al. 1989; Giorgi and Bates 1989). 
The finer regional-scale features of RCM can be attributed to detailed 
topography and land surface features, more comprehensive parameterization 
of unresolved physical processes  in the model equations, and explicit 
simulation of large mesoscale processes. 

 Atmosphere-ocean general circulation models (AOGCM) with the 
horizontal resolution of  a few hundred kilometers are currently used
for the  simulation of large-scale response of the climate system to
increasing of greenhouse gases and aerosol concentrations in the future.
The running of RCM  with the horizontal resolution of a few tens of
kilometers over an area of interest with boundary conditions of
AOGCM for the periods of 10-30 years in the  present and in the future 
can give additional information about the regional-scale climate and 
climate-change effects in this area.  Such downscaling studies related 
to climate change  have been made already for various parts of  Europe,
North America, Australia, and Africa; see for example the  references
cited by Jones et al. (1997), Laprise et al. (2003), Giorgi et al.
(2004), Duffy et al. (2006).   Currently some large projects such as
(PRUDENCE (Christensen et al. 2002) and NARCCAP
(http://www.narccap.ucar.edu)), launched  to investigate uncertainties in
the RCM climate-change simulations  over Europe and North America, are
underway. Multiple regional climate model ensembles are  used in these
studies in order to minimize uncertainties obtained in simulations with
each model.

The downscaling studies related to climate change  over South America
are just started.  The project "Climate change scenarios using PRECIS" 
(Jones et al. 2004) was launched by Hadley Center for Climate Prediction
and Research to develop user-friendly RCM which can be easily running on
personal computer for any area of the globe. The South American
countries including Brazil  are participated in this project running
PRECIS over various parts  of South America. The data of the atmospheric
global model HadAM3P were  provided by Hadley Center for using them  as
boundary conditions in these simulations.    The first regional
climatology for South America is presented by  Seth et al. (2007)  for
the period from 1982 to 2003. It was obtained by using the  RegCM3 model
(Pal et al. 2006) which was  forced both  by the reanalysis data 
(Kanamitsu et al. 2002) and by the European-Hamburg (ECHAM) AGCM 
(Roeckner et al. 1996)  global model.

The aim of this study is to propose one more regional climate model for
use in the climate downscaling research over South America.  For this aim 
we prepared  the climate version from the NCEP Eta regional forecast
model (Black 1994).  The Eta Model was chosen because it was
intensively used  for  weather forecast  as well as for seasonal
predictability and processes studies over South America during last
decade (Figueroa et al. 1995; Tanajura 1996;  Chou et al. 2000; 
Chou et al. 2002; Vernekar et al. 2003; Tarasova et al. 2006). 
Analysis of the integration results in most cases  demonstrates better 
agreement with observations of meteorological fields simulated by the 
Eta model as compared with AGCM. Nevertheless, the longest integrations 
with the Eta model were limited to the continuous integrations for 
3-5 months because of  the
limitations in the codes of the Eta model which was developed for the
forecast studies.   The climate versions of the Eta model which permit
integrations for the  period of any duration were developed at the
Brazilian Instituto Nacional de Pesquisas Espaciais/  Centro de Previsao
de Tempo e Estudos Cimaticos  (INPE/CPTEC) during last years 
(Pisnichenko et al. 2006;  Fernandez et al. 2006; Tarasova et al. 2006).

In order to be considered as a valid tool for dynamical downscaling
of low-resolution GCM fields a regional climate model has to
satisfy  some requirements (e.g., Wang et al. 2004;
Castro et al. 2005; Laprise, 2006).  Firstly, it is needed
to show that RCM is able to reproduce the principal  features  of
the large-scale fields of GCM which data is used
as driving  boundary  conditions and its main statistics.
It is necessary condition indicating that nonlinear interactions of 
small-scale components do not  strongly divert the system from 
the background state. This will also
guarantee that boundary conditions do not transform into
peculiarities. This is an issue of evaluation of consistency between
RCM and GCM  fields.  Secondly, it is necessary to show that RCM
adds small-scale features absent in the GCM driving fields and that
these features agree with observations and with high-resolution GCM
fields. Laprise et al. (2007) provide  a summary of studies related to this
issue. On the opinion of Laprise et al. (2007) the consensus on the
first point  is not yet reached within the RCMs community. It is not clear 
if  the large scales of GCM are unaffected, improved or
degraded by RCMs. Note, that a comparison of large-scale fields of RCM and 
GCM is mainly performed for the surface temperature and precipitation 
(e.g., Hudson and Jones  2002; Seth et al. 2007). Another type of comparison 
is presented by Castro et al. (2005) for the one month simulation of Regional 
Atmospheric Modeling System RAMS (Pielke et al. 1992) with the boundary 
conditions of the reanalysis. They did the spectral analysis of the column 
average  total kinetic energy and the column integrated moisture flux 
convergence and concluded from it that RAMS does not add increased skill  to 
the large scale available in the reanalysis.

In this work  we show the first results of our validation program related to  
the development of climate version of the Eta model. We investigated the consistency
of the large-scale output  fields of the Eta model and HadAM3P. For this, 
the geopotential, temperature  and wind fields  at various
levels were analysed by using Fourier analysis of time series,
consistency index, constituted  from linear regression coefficients,
time mean and space mean models' arithmetic difference (MAD), root mean square 
difference (RMSD), dispersion analysis and some others characteristics.  The short 
description of the Eta model and  the modifications, which we implemented in it,
is given in Section 2.  In this section the model integration
procedure is also described.  The newly developed version of the Eta
model is hereafter termed as INPE Eta for Climate Change Simulations
(INPE Eta CCS).  Section 3 presents the results of the integrations
with the INPE Eta CCS model over South America  driven by boundary
conditions from the HadAM3P for the period 1961-1991. The Eta model
output fields are compared with those from  HadAM3P in order to
prove a consistency between the two models. Section 4 gives summary
of the results and the conclusions.


\noindent {\bf 2. Model and experimental design}

For this work, aimed to prepare Eta model version for climate-change simulations, 
we initially adopted the workstation (WS) Eta modeling package 
(version of 2003) developed at the Science Operations Officer/Science and 
Training Resource Center (SOO/STRC). This package and its User Guide written 
by R. Rozumalski is freely available at http://strc.comet.ucar.  
The SOO/STRC WS Eta is nearly identical to WS 
Eta model and operational Eta Model of 2003, both developed at NCEP. 
Only the run-time scripts and model files organization 
were changed. The additional convection cumulus scheme 
of Kain and Fritsch (1993) was also implemented. 
The longest continuous integration with this model can be made 
for 1 month due to the restriction on the output file name, restart
subroutines, and some other impediments.   

\noindent {\it a. Short description of NCEP Eta model}

The full description of the NCEP Eta regional forecasting  model is given by  Mesinger
et al. (1988), Janjic (1994), and Black (1994).  In short, the horizontal field structure is
described  on a semi-staggered E grid.
The eta  vertical coordinate is used to reduce numerical errors over mountains in  
computing the pressure gradient  force. The  planetary boundary 
layer processes are described by the Mellor-Yamada  level 2.5 model 
(Mellor and Yamada 1974). The convective precipitation scheme is of Betts and
Miller (1986) modified by Janjic(1994). The shortwave  and longwave radiation  codes 
follow parameterizations of Lacis and Hansen (1974) and  Fels and Schwartzkopf (1975), 
respectively. The land-surface scheme is of Chen et al. (1997).  The grid-scale 
cloud cover fraction is  parameterized as a function of relative humidity and cloud 
water (ice) mixing ratio (Xu and Randall 1996; Hong et al. 1998). Convective 
cloud cover fraction is parameterized as a function of precipitation rate (Slingo 1987).

\noindent {\it b. Modifications in the SOO/STRC WS Eta model}

The SOO/STRC WS Eta model has been installed at supercomputer NEC SX6 at 
CPTEC. To be able to perform long term climate  integrations we have made 
multiple changes and corrections in the scripts and source codes of the original
model as well as have written the new programs. 

As it was already mentioned, the Eta model was forced at its  lateral and
bottom boundary by the output of HadAM3P model.  The  HadAM3P output data
represent    horizontal wind, potential temperature, specific humidity and 
earth surface pressure which are given on the horizontal Arakawa B-grid  and at
the 19 sigma-hybrid levels. These data are written in the PP-format. To use them 
for the Eta model boundary conditions these data have to be transformed  
into horizontal wind, geopotential, mixture ratio and  earth surface pressure 
given on  regular latitude-longitude grid at standard p-surface levels. 
For this aim, some of the pre-processing Eta model programs were modified and 
new program which converts the HadAM3P output data to those acceptable by 
the Eta model was written.

Another modifications made in the Eta model can be shortly described  as
following. There were re-written the  SST update programs used to accept  the
SST and SICE data generated by HadCM3 every 15 days.  The programs of the Sun's 
elevation angle and of calendar were modified in order to be able to integrate the Eta
model for the artificial year of 360 days which is used by HadAM3P. There were
developed new restart programs which can be used in multiprocessing
integration. These programs allow to continue the model integration   from any
time moment  by using the model output binary files.  This is the useful  option
for long term climate integration because of the large size of the   file of
boundary conditions needed for continuous integrations.  Another reason for
use of the restart option is the large size of the output binary files which
after post-processing can be written in more economic GRIB format.   All
shortcomings which restrict a period of model integration  were
corrected including those in the post-processing subroutines.

The additional solar radiation scheme (CLIRAD-SW-M)  developed by  Chou and
Suarez (1999) and modified by Tarasova and Fomin (2000)  was implemented in the
model. The results  of the month integration with this scheme were analysed 
by Tarasova et
al. (2006). The additional thermal radiation scheme  of Chou et al. (2001) was
also implemented. This allows to run the model with increasing concentration of
${\rm CO_2}$ and other trace gases needed for  future climate simulation
experiments. All these corrections, modifications and implementations were 
made  taking into account that the model can  be run on Linux cluster or any
other multi-processors computer.

\noindent {\it c. Integration with the INPE Eta CCS model}

The first step in evaluation of dynamical downscaling results is
investigation of a consistency between 
regional model outputs and GCM data used for RCM boundary conditions. That
is, we have to show that our RCM does not significantly diverge from GSM in
reproducing time mean large scale patterns of circulation. We also expect that
both models reproduce a low-frequency oscillation of the atmosphere in a similar
manner. 

For this aim we analysed the results of the Eta CCS model integration for the period 
1960-1990 over South America.  These data are the part of the results  of current and
future  climate
downscaling  experiments covering the periods of 1960-1990 and 2071-2100, respectively. 
The detailed analysis of the results of these experiments is currently making 
by our group and will  be present in further publications.

The Eta CCS model in our experiments was forced at its lateral and bottom boundary by
the output of HadAM3P, which was run using  SST, SICE (sea ice) and
greenhouse  gases and aerosol concentration as external driving from coupling
model HadCM3.  Data for lateral boundary conditions for the Eta CCS model were provided 
every 6 hours and SST and SICE data every 15 days. Linear interpolation for values 
on lateral boundaries, SST, and SICE was used between these periods. For the
initial conditions of soil moisture and soil temperature the climate mean 
values were used. The spin up period of soil moisture and temperature we have 
accepted  to be equal to 1 year. Hence, the first year of the integration was 
not used in the analysis.

The area of the integration was centered at  $58.5^{\circ}$ W longitude and 
$22.0^{\circ}$ S latitude  and covers the territory of South American continent
with adjacent oceans ($55^{\circ}$ S - $16^{\circ}$ N, $89^{\circ}$ W -
$29^{\circ}$ W).  The model was integrated on the 211$\times$115 horizontal grid with
grid spacing of 37 km. In the vertical, 38 eta coordinate layers were  used. For
the modern climate integration the Betts-Miller cumulus convection parametrization
scheme and the ETA model original shortwave and longwave radiation schemes
were chosen. 

\bigskip
\bigskip
\bigskip

\noindent {\bf 3. Analysis of the integration results}

The verification of a consistency between the outputs of the Eta CCS model and HadAM3P 
is particularly important due to the difference between the physical parameterization 
packages of these two models. 
To prove an agreement between these models results we have  compared the geopotential height, 
temperature and kinetic  energy fields on the earth surface and
at the various p-levels (1000 mb, 700 mb, 500 mb) from these two data sources.
More detailed comparison was made for the five regions shown in Figure 1: 
Amazonia ($12.5^{\circ}$ S - $5^{\circ}$ N,  $75^{\circ}$ W - $48.75^{\circ}$ W); Nordeste
(north-east of Brazil) ($15^{\circ}$ S - $2.5^{\circ}$ S, $45^{\circ}$ W - $33.75^{\circ}$
W); South of Brazil ($32.5^{\circ}$ S - $22.5^{\circ}$ S, $60^{\circ}$ W - $48.75^{\circ}$
W); Minas ($22.5^{\circ}$ S - $15^{\circ}$ S, $48.75^{\circ}$ W - $41.25^{\circ}$ W);
Pantanal ($17.5^{\circ}$ S - $12.5^{\circ}$ S, $60^{\circ}$ W - $52.5^{\circ}$ W).
The time averaged fields and time series of space averaged meteorological
variables were analysed.

\noindent {\it a. Methods of the analysis}

A number of measures of consistency between the  
outputs of the Eta CCS regional model (hereafter RM) and HadAM3P global model 
(hereafter GM) are used here. The original package of programs was developed 
for comparing the models. 
First, we assessed the climatological means and time averaged difference
between the models, which give an opportunity to identify systematic differences 
between the models. Then we analysed various characteristics which allow to show 
in detail a distinction between GM and RM simulated fields. Since this work is dedicated
to investigation RM abilities to reproduce mean fields of driving GM and some their 
statistical moments therefore the regional model fields were scaled to the global model 
grid. For this aim we removed the small scale component from the regional model 
fields applying smoothed filter. This filter is the two dimensional version 
of the weighted moving averages, where the weights depend linearly on the 
distance between the grid point of the global model and the grid points of the 
regional model (in which are sited the data used in smoothing procedure). 
The weight increases when the distance decreases. This smoothing procedure can be 
written as:
 
\beq
  \Phi(x_i,y_j)=\sum_{r_{i,j;k} < r_0} \phi(\hat{x}_k,\hat{y}_k)\, p_k 
\eeq

where $\Phi(x_i,y_j)$ is
a smoothed value of regional model field on global grid point, $ r_0$ is the radius 
of influence which defines the circle inside which the RM field data are used for average
calculation, $r_{i,j;k}$ - the distance from a $(x_i,y_j)$ point of GM grid to $k$-th 
RM grid point $(\hat{x}_k,\hat{y}_k)$,  $\phi(x_k,y_k)$ are the field value at $k$-th 
RM grid point inside the circle defined by the radius of influence, $p_k$ is a weight 
for the $k$-th RM grid point and which is calculated as
 
\beq 
p_k \,=\, \left (1-\frac{r_{i,j;k}}{r_0}\right ) / \left( \sum_{r_{i,j;k} < r_0}1 - \frac{1}{r_0}
\sum_{r_{i,j;k} < r_0} r_{i,j;k}\right ). 
\eeq

In this formula the numerator decreases with increasing $r_{i,j;k}$ and becomes equal 
to zero when $r_{i,j;k}$ is equal or larger than $r_0$. The denominator is defined 
from a normalization condition, namely a sum of all $p_k$ weights must be equal to $1$. 
  
In order to compare the models we analysed how they reproduce the time average fields 
of meteorological variables as well as the fields of standard deviation of these variables. 
For more detailed assessment of the consistency between the RM and GM fields 
we also calculated the models' arithmetic difference and coefficients of linear 
regression using time-series of meteorological variables at each grid point of 
the Eta model. The fields of these characteristics present useful information about 
a degree of consistency of the models results. 

For quantitative and direct description of the consistency between the RM and GM 
output fields  we propose to use a new characteristic which we termed a consistency 
index (CI).  This characteristic represents  some integral variant of Taylor 
diagram (Taylor, 2001). In terms of correlation coefficient, standard deviations, 
and mean values of compared fields, it expresses the resemblance of one field to another. 
It is some simple functional which depends on coefficients of linear regression 
of one field on another field. We found usefulness of this characteristic in the 
capability to describe the similarity of two fields by one number only in the case 
when the space patterns are analysed. The use of unique number for describing 
the resemblance of two random series is of particular interest in the case  
when an analysis of consistency of the time evolution of the space patterns is
performed. We can analyse in this case the time  series of compared fields at every 
grid point and describe the resemblance of the  time evolution of analysed fields 
by one field only  (namely, the consistency index number at every grid point).

The numeric value of CI we define as

\beq
      CI = 
\begin{cases}
\displaystyle (1 - \frac{\Delta S_d}{\Delta S_n} ) \frac{\sigma_G}{\sigma_R}
 \, \, \textrm{ for } \frac{\sigma_G}{\sigma_R}  \le 1\textrm{,}\\ 
\displaystyle (1 - \frac{\Delta S_d}{\Delta S_n} )\frac{\sigma_R}{\sigma_G}
 \, \, \textrm { for } \frac{\sigma_G}{\sigma_R}  > 1. 
\end{cases}
\eeq

Here $\sigma_G$ and $\sigma_R$ are the sample standard deviation of investigated 
meteorological parameter of a global model series and a regional model series, 
respectively. The $\Delta S_d$ is the area of figure $ABOCD$ (see Figure 2) 
which is  formed by two straight lines of linear regression and two verticals which
intersects them. The straight line $r$ is a linear regression line of the GM series
on RM series. The straight line $i$ is an ideal regression line 
for the identical GM and RM serieses with regression coefficients $a0=0$ and $a1=1$.
The two verticals that intersect these regression lines have the coordinates of
$x_R=a-s$ and $x_R=a+s$. The $a$ is a mean  value of investigated meteorological 
parameter of the RM series normalized on  $s_0=1.44 \sigma_R$. The $s$ is a
nondimensional  value of $s_0$. The interval ($a-s$, $a+s$) contains  $85\%$ of 
members of the RM series (under the assumption  that the series obeys the Gaussian 
distribution). $\Delta S_n$ is the area  of a triangle $BCE$. The area of the shaded 
figure $ABOCD$  statistically describes a degree of resemblance of the GM and RM 
serieses: Smaller area corresponds to closer resemblance. The area of the triangle  
$BCE$ is equal to $2$ in nondimensional coordinates  and  describes the case when 
the RM and GM serieses are non-correlated and the mean value of the GM series is 
equal to $a - s $  (or $a + s$). The multiplier  $\frac{\sigma_G}{\sigma_R}$
($\frac{\sigma_R}{\sigma_G}$ ) approximately describes the ratio of transient-eddy
amplitudes reproduced by the models under comparison. Ideally, these amplitudes must 
be very close. The magnitude of CI is close to $1$ if the GM and RM series  
statistically resemble one another and it is equal to zero or to  negative value 
when there is no similarity  of the serieses.  When $ABOCD$ is larger than
$BCE$ the CI is less than zero what means that the resemblance of the serieses is 
worse than for the  non-correlated serieses with the mean value of the GM series 
larger (or smaller) than $a + s$ ($a - s$).

Since we had to process very large amount of data, we used recurrence formulas 
for the calculation  of averages,  sample standard deviations, and coefficients of 
linear  regression for various GM and RM serieses and wrote these characteristics 
to the model output  every 24 hours. These characteristics for any time period can be 
recalculated from this running statistics. The recurrence formulas  and formulas 
that were used for recalculation are presented in appendix A.

\noindent {\it b. Assessment of the RM and GM consistency}

At first we present geopotential height, temperature and kinetic energy 
fields averaged over the period of integration from 1961 to 1990. Figures 3 and 4 
show these fields at the  levels of 1000 mb and 700 mb, respectively, obtained from the 
RM and GM integrations. A comparison of both models fields at the 1000 mb level  
shows good agreement between the fields of geopotential height and between 
the temperature fields.  There is  general agreement between 
the kinetic energy fields. Some disagreement in the temperature magnitude exists 
in the central part of tropical South America.  
The values of kinetic energy differ over most part of the continent. This is probably 
related  to the different physical parameterization packages in these models.
The same RM and GM fields at the higher level of 700 mb bear closer spatial and 
quantitative resemblance. Note, that the fields similarity at 500 mb (not shown) is 
higher than that at 700 mb. This is a consequence of the diminishing of the impact of 
surface-atmosphere interaction on the higher-level atmospheric circulation. We also
compared the same RM and GM fields averaged over January and July (not shown). The
agreement between the fields is better in July (austral winter) than in January (austral
summer). The fields of time standard deviation of meteorological variables provide additional 
information about an amplitude of their temporal fluctuations.  Figure 5 and 6 presents 
the RM and GM standard deviation fields of  geopotential height, temperature and kinetic 
energy at the 1000 mb and 700 mb levels averaged over the period of integration. One can  
see reasonably high degree of consistency between the  RM and GM 
standard deviation fields. The standard deviation fields also bear closer resemblance 
for geopotential height and temperature than for kinetic energy. With the increase of 
altitude the difference between the RM and GM standard deviation fields is diminished for 
all variables.

The quantitative distinction between the two fields is usually described by the 
field of models' arithmetic difference (MAD) that is the difference between the fields 
values at each grid point. The left column of Figure 7 shows MAD between 
the RM and GM  geopotential height, temperature, and kinetic energy fields at 1000 mb 
averaged over the period of integration.  One can see that the largest values of this
field are seen over the tropical and sub-tropical parts of the Southern American 
continent. The significant values of MAD over the Andes is probably related to the errors 
of interpolation from the sigma-hybrid surfaces to the pressure surfaces located below 
the Earth's surface in the global model.  With increasing of the altitude (700 mb, 500 mb) 
the values  of MAD decrease for all fields (Figure 8).  The MAD of these variables 
(geopotential height, temperature, and kinetic energy) averaged over July (January) 
is smaller (larger) than that averaged over all period of integration. 

The right column of Figure 7  presents the consistency index (CI) fields for geopotential 
height, temperature, and kinetic energy at the level of 1000 mb. The magnitude of CI which 
is close to 1 means good resemblance between the RM and GM fields. The CI fields resemble 
the fields of MAD in terms of spatial distribution. But the use of non-dimensional  CI 
allows to compare quantitatively a similarity of the fields of different meteorological 
variables. Thus, the CI fields in Figure 7 show that the consistency of the fields of 
geopotential height is higher than that of the temperature fields and the consistency of 
the kinetic  energy field is lower than that of both geopotential height and temperature.

To compare the model outputs we also analysed a temporal variations of the geopotential
height, temperature and kinetic energy values at 1000 mb and 500 mb levels,   averaged
over all integration domain and over the regions shown in Figure 1. Figure 9 presents 
monthly mean models' arithmetic difference and root mean square difference  (RMSD) between 
the GM and RM time serieses for these variables averaged over the integration domain.
For each variable  the upper figure represents MAD and the lower figure shows RMSD. One 
can see that the magnitude of mean MAD is not high. It is about  6 m   in geopotential 
height,   less than 0.1 $^{\circ}$K in temperature, and about 10 m$^{2}$\,sec$^{-2}$  
in kinetic energy at 1000 mb. The mean RMSD values at 1000 mb are not high also. Its 
magnitude is  about 24 m in geopotential heights,  3.4 $^{\circ}$K in temperature, and 
39 m$^{2}$\,sec$^{-2}$ in kinetic energy. Low  magnitude of RMSD proves that current 
absolute values of MAD are not high for each moment of integration. Figure 9 shows also 
that there is no drift of MAD and RMSD during the integration that proves the RM 
integration stability. The magnitude of temporal correlation coefficient  between the time 
serieses of the RM and GM space averaged fields is about 0.95-0.98. This means that RM 
follows the GM boundary driving.  At the level of 500 mb as MAD as RMSD  are of lower or 
same  magnitude.   We also analysed the same time series for the above  mentioned regions 
(Amazonia, Nordeste, South of Brazil, Minas, Pantanal). The correlation coefficients 
between the RM and GM time serieses as well as mean MAD and RMSD at 1000 mb and 500 mb 
are shown in Table 1 for all domain and for the five regions. 
One can see that these coefficients slightly varies from region to region. Note one case
of low correlation between the kinetic energy time series at 1000 mb in Amazonia related to
low magnitude of wind at the surface level in GM.

Figures 10 and 11 show the time evolution of annual mean MAD in the geopotential height,
temperature and kinetic energy fields at 1000 mb and 700
mb, respectively, for the above mentioned regions. At the 1000 mb level 
the magnitude of MAD for different regions varies  from -10 m to +17 m for geopotential 
height, from  -4.0$^{\circ}$K to +0.3$^{\circ}$K for temperature, and  from 
-20 m$^{2}$\,sec$^{-1}$ to -5 m$^{2}$\,sec$^{-1}$ for kinetic energy. The amplitude 
of interannual variations of these meteorological variables differs from one region to
another.  We can see that there is no significant trend  and strong fluctuations 
of MAD for any region.  A significant mutual correlation between the MAD obtained 
for various regions does not exist. Note that the values of MAD and 
the amplitudes of its  interannual variations for geopotential height 
and temperature decrease when the altitude  increases. For kinetic energy  both MAD and
amplitude of interannual variations increase when the altitude increases. Though the
magnitude of relative MAD (for example, that divided by a mean standard deviation) 
for kinetic energy also decreases. 

Figure 12 presents  a scattering diagram of daily  linear regression coefficients values
(a0, a1) which describe the regression of the GM 1000 mb  geopotential height field on the same
RM field (top);  time evolution of these linear regression coefficients (a0 , a1) (middle)
for each month of the model run; and the time evolution of consistency index (bottom). 
The consistency index was calculated in the same way as described above (Figure 2), but the time
series were substituted by "space" series formed by variable values at all grid points. 

Concerning this  figure we can say that in the hypothetical case, when the fields of GM and
RM coincide, all points in the top figure will fall on one point with the coordinates a1=1.0
and a0=0.0.   Thus we can affirm  that if the points on the top figure are 
located near the  point (a1=1, a0=0) the RM and GM fields  are  very similar; in the case 
when the points are reasonably scattered  but the center of mass  of this distribution is 
close to the point (a1=1, a0=0) we can say that the fields of the models  are similar in 
average. The time series of linear regression coefficients a0 and a1 of GM data  upon RM 
data have large negative correlation (middle figure). In the most cases it leads to some 
compensation in the variations of CI shown on the bottom figure. The CI variations clearly 
express the year oscillation. Its mean value is about 0.94  and increases with the altitude. 
Its linear time trend is very small. This provides some more indication that the considered 
models do not diverge. Figure 13 presents the same  characteristics  as shown in Figure 12
but for the RM and GM temperature fields at 1000 mb. The scattering diagrams in this case 
indicates  that GM is  slightly warmer then RM for the regions with low
temperatures and slightly colder  for the regions with higher temperatures. This is in
agreement with Figure 3 which shows mean temperature fields for all period of the
integration.

For more detailed analysis of the time evolution of mean values of  meteorological
variable fields  we have calculated spectral distribution of their time  series  by using 
Fast Fourier  Transform  algorithm. Figure 14 shows an example of such distribution for 
the time series of geopotential height, temperature and kinetic energy averaged over all 
integration domain. One can see that the GM and RM spectras  have a high degree of 
similarity. The high frequency tails quasi coincide. The year and semi-year oscillations 
have the same amplitude. Four year cycle in geopotential height and temperature is 
reproduced by RM and GM quasi identically. This cycle in kinetic energy spectra is also 
reproduced by both models but not identically. Also the models agree in reproducing of 
6-9 years  minimum and of the next increase of the spectra. Quasi all synoptic and seasonal 
oscillation maximums coincide in the RM and GM spectras. We calculated the same spectras 
for above mentioned regions shown in Figure 1. The RM and GM spectras for these regions 
demonstrate similar coincidence as that for all integration domain with insignificant 
distinctions. Only for the Pantanal region, the spectras of GM and RM  kinetic energy  at
1000 mb diverge significantly. But with the increase of altitude  this difference
diminishes and quasi disappears at 500 mb.

\noindent {\bf 4. Conclusions}  

This analysis of the output results of 30-year runs of regional model and 
its driving global model  confirms that the models have a high degree of 
consistency despite of the difference in their physical parameterizations. 
In the future work we are planning  to estimate an  
impact of tuning in RM physical parameterizations such as radiation and 
convection schemes on consistency of RM and GM output fields.  
An impact of the use of another driven global model on the RM and GM resemblance 
will be  also estimated. We also need to evaluate the model performance 
for current climate by comparing regional model outputs with observations on global 
and regional scales. In order to estimate the impact of global model errors
on the regional model outputs, the integration of the regional model driven 
by the reanalysis data (Kanamitsu et al. 2002) is planned. 
The approach developed in this paper can form the basis for quantitative assessment 
of regional model and its driving global model consistency. Currently,
many researchers use various regional models for dynamical downscaling but a few
publications exist about the quantitative assessment of the similarity
between the large-scale fields of a regional model and its driving  global
model. 
Even if regional and global models have the same physical
parameterization packages, the difference between the models can be related to
the low time frequency and low space resolution of boundary forcing 
in the regional model.
 
\appendix

\noindent {\bf Appendix A } 
\noindent {\it Recurrence formulas }

For the evaluation of the consistency of the models we analysed very large serieses 
of the meteorological data. To make the work with series faster and for 
economy of computer resources we used recurrence formulas for calculating
running average, standard deviation and covariance, from which we can calculate
any others necessary characteristics. 

We accept the definition of running mean, variance and covariance 
respectively as

\beq 
\tag{A1}
\label{eq:mea}
\bar{x}_n \, = \, \frac{1}{n} \sum_{i=1}^n x_i ,
\eeq
 
\beq
\tag{A2}
\label{eq:var}
D_n \, =\, \frac{1}{n} \sum_{i=1}^n (x_i - \bar{x}_n)^2 ,
\eeq

\beq
\tag{A3}
\label{eq:covar}
r_n \, =\,  \frac{1}{n} \sum_{i=1}^n (x_i - \bar{x}_n)(y_i - \bar{y}_n).
\eeq 
Here  $\bar{x}_n$, $D_n$, and  $r_n$ are the sample mean, the  sample variance, 
and the sample covariance for serieses containing  $n$ terms, $x_i$, $y_i$ are 
the i-th term of series.  
The recurrence formula for a sample mean is obvious

\beq
\tag{A4}
\label{eq:avrec}
\bar{x}_n \, =\, \frac{n-1}{n}\bar{x}_{n-1} + \frac{1}{n}x_n. 
\eeq

Below we derive the recurrence formula for a sample covariance. The analogous formula 
for a sample variance is obtained after replacing  $y_i$, $\bar{y}_n$ by 
$x_i$,  $\bar{x}_n$.

Let us rewrite formula (\ref{eq:covar}) using (\ref{eq:avrec}) in following manner

\bd
\begin{split}
&r_n \, =\,  \frac{n-1}{n} \cdot \frac{1}{n-1} \sum_{i=1}^{n-1} (x_i - \frac{n-1}{n} 
\bar{x}_{n-1}-\frac{1}{n} x_n)(y_i - \frac{n-1}{n} \bar{y}_{n-1}-\frac{1}{n} y_n) + \\
&+ \frac{1}{n}(x_n - \bar{x}_n)(y_n - \bar{y}_n)
\end{split}
\ed

Now we group the members of this formula to select the part that is equal to the
covariance on previous $(n-1)$ step 
\bd
\begin{split}
&r_n \, =\,  \frac{n-1}{n} \cdot \frac{1}{n-1} \sum_{i=1}^{n-1} (x_i - \bar{x}_{n-1})
(y_i - \bar{y}_{n-1}) + \frac{1}{n} (\bar{y}_{n-1}-y_n) \cdot \frac{n-1}{n} \cdot 
\frac{1}{n-1} \sum_{i=1}^{n-1} (x_i - \\
&- \bar{x}_{n-1}) + \frac{1}{n} (\bar{x}_{n-1}-x_n) \frac{n-1}{n} \cdot \frac{1}{n-1} 
\sum_{i=1}^{n-1} (y_i - \bar{y}_{n-1}) + \frac{1}{n} (\bar{x}_{n-1}-x_n)\cdot 
\frac{1}{n} (\bar{y}_{n-1}-y_n) \cdot \\
& \cdot \frac{n-1}{n} + \frac{1}{n} (x_n - \bar{x}_n) (y_n - \bar{y}_n).
\end{split}
\ed

Taking into account that the terms $\frac{1}{n-1} \sum_{i=1}^{n-1} (x_i - 
\bar{x}_{n-1})$  and $\frac{1}{n-1} \sum_{i=1}^{n-1} (y_i - \bar{y}_{n-1})$ are equal
to zero  and using again  formula (\ref{eq:avrec}) we obtain
\beq
\tag{A5}
\label{eq:covrec}
r_n \, =\, \frac{n-1}{n} r_{n-1} + \frac{n-1}{n^2}(\bar{x}_{n-1}-x_n)
(\bar{y}_{n-1}-y_{n}) .
\eeq
 
Finally we show how to recalculate these running values for any time interval.
Let $\bar{x}_m$ be the mean value for series from the first $m$ elements of 
$x_i$ and let $m < n$. Denote  $\bar{x}_{m:n}$  the mean  value of $x_i$  
for the series $x_{m+1}, x_{m+2},... x_n $ as
$$
\bar{x}_{m:n} \, = \, \frac{1}{n-m} \sum_{i=m+1}^n x_i ,
$$

It is easy to obtain that 
\beq
\tag{A6}
\label{eq:avint}
\bar{x}_{m:n} \, =\, \frac{1}{n-m}(n \bar{x}_n - m \bar{x}_m).
\eeq

Now, let us derive formula for calculating the covariance for interval 
$(m+1, n)$ using the meanings for covariance and average for intervals 
$(1,m)$ and $(1,n)$.

\beq
\tag{A7}
\label{eq:covin1}
n \bar{r}_n - m \bar{r}_m \, =\, \sum_{i=1}^n(x_i y_i) -n \bar{x}_n \bar{y}_n -
 \sum_{i=1}^m(x_i y_i) + m \bar{x}_m \bar{y}_m 
\eeq 

Taking into account that 
\beq \tag{A8}
\label{eq:covintd}
(n-m) \bar{r}_{m:n} \,=\, \sum_{i=m+1}^n(x_i y_i) - (n-m)\bar{x}_{m:n}
\bar{y}_{m:n}, 
\eeq 
we rewrite (\ref{eq:covin1}) as
\beq
\tag{A9}
n \bar{r}_n - m \bar{r}_m \, =\, (n-m) \bar{r}_{m:n} -n \bar{x}_n \bar{y}_n
+ m \bar{x}_m \bar{y}_m + (n-m)\bar{x}_{m:n}\bar{y}_{m:n}.
\eeq

Lastly, substituting the $\bar{x}_{m:n},\bar{y}_{m:n}$ 
from formula (\ref{eq:avint}) and making routine transformations we obtain
the desired formula
\beq
\tag{A10}
\bar{r}_{m:n} \, =\,  \frac{1}{n-m}(n \bar{r}_n - m \bar{r}_m) -
\frac{mn}{(n-m)^2}(\bar{x}_n - \bar{x}_m)(\bar{y}_n - \bar{y}_m).
\eeq

{\it Acknowledgments.} I.A. Pisnichenko was supported by Global Opportunity 
Fund (GOF) from UK Foreign Commonwealth Office, T.A. Tarasova was sponsored 
by INPE/CPTEC as part of an international agreement with the NEC Corporation. 
The authors thank their colleagues from CPTEC/INPE  C. Nobre and J. Marengo 
for  their administrative contributions that made it possible 
for us to perform this work. The authors also thank Hadley Center for presenting
HadAM3P data.

\newpage 

\centerline{{\bf References }}

\setlength{\parindent}{-1.0\parindent}

Betts AK, and Miller MT (1986) A new convective adjustment 
scheme. Part II: Single column tests GATE wave, BOMEX, and 
Arctic air-mass data. Quart J Roy Met Soc 112: 693-703  

Black TL (1994) NMC notes: the new NMC mesoscale Eta model: 
description and forecast examples. Wea Forecasting 9:256-278

Castro CL, Pielke Sr RA, and Leoncini G (2005)
Dynamical downscaling: Assessment of value retained 
and adding using the Regional Atmosphering Modeling System (RAMS).
J Geophys Res 110(D05108) doi: 10.1029/2004JD004721

Chou M-D, and Suarez MJ  (1999) A solar radiation 
parameterization (CLIRAD-SW) for atmospheric studies. Preprint  
NASA/Goddard Space Flight Center, Greenbelt, Maryland, 38 pp

Chou M-D, Suarez MJ, Liang X-Z, and Yan M M-H (2001) A thermal 
infrared radiation parameterization for atmospheric Studies.
Preprint NASA/Goddard Space Flight Center, Greenbelt, Maryland, 55 pp
       
Chen FK, Janjic Z, and Mitchel K, (1997) Impact of the 
atmospheric surface-layer parameterizations in the new 
land-surface scheme of the NCEP mesoscale Eta model. Bound-Layer Meteor 
85: 391-421   

Christensen JH, Carter T, and Giorgi F (2002) PRUDENCE employs 
new methods to assess European climate change. EOS Trans Amer 
Geophys Union 82:147

Chou SC, Nunes AMB, and Cavalcanti IFA (2000) Extended forecast over 
South America using the regional Eta model. J Geophys Res 
105:10147-10160.

Chou SC, Tanajura CAS, Xue Y, and Nobre CA (2002) Validation of the
coupled Eta/SSiB model over South America. J Geophys Res 107:8088  
doi:10.1029/2000JD000270 

Duffy PB, Arritt RW, Coquard J, Gutowski W, Han J, Iorio J, Kim J,
Leung L-R, Roads J, Zeledon E (2006) Simulations of present and 
future climates in the western United States with four nested regional 
climate models. J Clim 19:873-895 

Dickinson RE, Errico RM, Giorgi F, and Bates GT (1989)  
A regional climate model for the western United States. 
Clim Change 15:383-422

Fels SB, and Schwartzkopf MD (1975) The simplified exchange 
approximation: A new method for radiative transfer 
calculations. J Atmos Sci 32:1475-1466

Fernandez JPR, Franchito SH, and Rao VB (2006) Simulation 
of the summer circulation over South America by two regional climate models.
Part I: Mean climatology. Theor Appl Climatol 86:247-260

Figueroa SN, Satyamurty P, and de Silva Dias PL (1995) Simulations
of the summer circulation over the South American region with the
Eta coordinate model. J Atmos Sci 52:1573-1584 

Giorgi F, and Bates GT (1989) The climatological skill of a regional 
model over complex terrain. Mon Wea Rev 117:2325-2347

Giorgi F, Bi X, Pal JS (2004) Mean, interannual variability and trends 
in a regional climate change experiment over Europe. I. Present-day 
climate (1961-1990). Clim Dyn 22:733-756 doi: 10.1007/s00382-004-0409-x 

Hong S-Y, Yuang H-M, and Zhao Q (1998) Implementing of 
prognostic cloud  scheme for a regional spectral model. Mon Wea Rev 
126:2621-2639

Hudson DA, and Jones RD (2002) Regional climate model simulations of present-day
and future climates of Southern Africa. Hadley Centre for Climate Prediction and
Research, Met Office, Bracknell, UK

Janjic ZI (1994) The step-mountain eta coordinate model: further 
development of the convection, viscous sublayer, and 
turbulence closure schemes. Mon Wea Rev 122:927-945

Jones RG, Murphy JM, Noguer M, Keen AB (1997) Simulation of climate 
change over Europe using a nested regional-climate model. II: Comparison of 
driving and regional model responses to a doubling of carbon dioxide. 
Quart J Roy Met Soc 123:265-292

Jones RG, Noguer M, Hassel DC, Hudson D, Wilson SS, Jenkins GJ,
and Mitchell JFB (2004) Generating high resolution climate change scenarios
using PRECIS. Met Office, Hadley Center, Exeter, UK, 40 pp

Kanamitsu M, Ebisuzaki W, Woollen J, Yang S-K, Hnilo JJ, 
Fiorino M, and Potter GL (2002) NCEP-DOE AMIP-II Reanalysis 
(R-2). Bull Amer Meteor Soc 83:1631-1643

Kain JS, and Fritsch JM (1993) A one-dimensional entraining detraining 
plume model and its applications in convective parameterization.
J Atmos Sci 23:2784-2802 

Lacis AA, and Hansen JE (1974) A parameterization for the 
absorption of solar radiation in the Earth's atmosphere. J Atmos Sci 
31:118-133

Laprise R, Caya D, Frigon A, Paquin D (2003) Current and perturbed climate 
as simulated by the second-generation Canadian Regional Climate Model 
(CRCM-II) over north-western North America. Clim Dyn 21:405-421 
doi: 10.1007/s00382-003-0342-4 

Laprise R (2006) Regional climate modelling. 
J Comput Phys doi: 10.1016/j.jcp.2006.10.024

Laprise R, de Elia R, Caya D et al (2007) Challenging some tenets of 
regional climate modelling. 

Mellor GL, and Yamada T (1974) A hierarchy of turbulence 
closure models for boundary layers. J Atmos Sci 31:1791-1806.	

Mesinger F, Janjic ZI, Nickovic S, Gavrilov D, and Deaven DG (1988) The
step-mountain coordinate: model description and performance for cases of
Alpine lee cyclogenesis and for a case of Appalachian redevelopment. Mon
Wea Rev 116:1493-1518 

Roeckner E, Aroe K, Bengtsson L et al (1996) The atmospheric general 
circulation model ECHAM-4: model description and simulation of 
present day climate. Technical report, 218, Max-Plank Institute for 
Meteorology

Pal JS, Giorgi F, Bi X et al (2006) The ICTP RegCM3 and RegCNET: 
regional climate modeling for the developing World. 
Bull Am Meteorol Soc (in press)

Pielke Sr RA et al (1992) A comprehensive meteorological 
modeling system - RAMS. Meteorol Atmos Phys 49:69-91

Pisnichenko IA, Tarasova TA, Fernandez JPR, and 
Marengo J (2006) Validation of the Eta WS regional climate model
driven by boundary  conditions from the HADAM3P over South America. 
Proceedings of 8 ICSHMO, Foz do Iguacu, Brazil, April 24-28,  
INPE, 595-597

Seth A, Rauscher SA, Camargo SJ (2007) RegCM3 regional climatologies
for South America using reanalysis and ECHAM global model driving fields.
Clim Dyn 28:461-480 doi: 10.1007/s00382-006-0191-z

Slingo JM (1987) The development of a cloud prediction model 
for the ECMWF model. Quart J Royal Met Soc 113:899-927 

Tanajura CAS (1996) Modeling and analysis of the South American
summer climate. PhD Thesis, Univ of Md, College Park, 164 pp

Tarasova TA, and Fomin BA (2000) Solar radiation absorption 
due to water vapor: Advanced broadband parameterizations. J Appl Meteor 
39:1947-1951

Taylor KE (2001) Summarizing multiple aspects of model performance in single
diagram:  J Geophys Res 106:7183-7192

Vernekar AD, Kirtman BP, Fennessy MJ (2003) Low-level jets and their effects 
on the South American summer climate by the NCEP Eta Model. J Clim 16:297-311 

Tarasova TA, Fernandez JPR, Pisnichenko IA, Marengo JA,
Ceballos JC, and Bottino MJ (2006) Impact of new solar radiation
parameterization in the Eta Model on the simulation of summer  climate over
South America. J Appl Meteor Climatol 44:318-333 

Wang Y, Leung LR, McGregor JL, Lee D-K, Wang W-C, Ding Y, and Kimura F (2004) 
Regional climate modeling: progress, challenges, and prospects. J Meteor Soc Jpn
82:1599-1628 

Xu K-M, and Randall DA (1996) A semi empirical cloudiness
parameterization  for use in climate models. J Atmos Sci 
53:3084-3102

\newpage 

\centerline{{\bf Figure captions}} 

\bigskip

\noindent {\bf Figure~1.} The regions over South America selected for the analysis:
Amazonia (1), Nordeste (2), Sul Brasil (3), Minas (4), 
Pantanal (5).

\noindent {\bf Figure~2.} Definition of consistency index by using
the coefficients of  linear regression of HadAM3P field on Eta CCS model field. 

\noindent {\bf Figure~3.} Mean (1961-1990) fields of geopotential height
(m), temperature ($^{\circ}$K), and kinetic energy (m$^{2}$\,sec$^{-2}$) at 1000 mb, 
provided by HadAM3P (left) and Eta CCS model (right) simulations.  

\noindent {\bf Figure~4.} The same as in Figure 2 but at 700 mb.

\noindent {\bf Figure~5.} Mean (1961-1990) standard deviation fields of geopotential 
height (m), temperature ($^{\circ}$K), and kinetic energy (m$^{2}$\,sec$^{-2}$) 
at 1000 mb, provided by HadAM3P (left) and Eta CCS model (right) simulations.

\noindent {\bf Figure~6.} The same as in Figure 4 but at 700 mb. 

\noindent {\bf Figure~7.} Mean (1961-1990) fields  of MAD (left), 
calculated for HadAM3P and Eta CCS model fields of  
geopotential height (m), temperature ($^{\circ}$K), and 
kinetic energy (m$^{2}$\,sec$^{-2}$) at 1000 mb, and consistency index between
HadAM3P and Eta CCS model(right), calculated for the same fields.
 
\noindent {\bf Figure~8.} The same as in Figure 7 but at 700 mb. 

\noindent {\bf Figure~9.} Time series of mean (over the integration domain) 
MAD and RMSD, calculated for HadAM3P and Eta CCS model fields of geopotential height (m), 
temperature ($^{\circ}$K), and kinetic energy (m$^{2}$\,sec$^{-2}$) 
at 1000 mb (left) and 500 mb (right).

\noindent {\bf Figure~10.} Time series of mean (over the regions shown in Figure 1)
MAD, calculated for HadAM3P and Eta CCS model fields of geopotential height, G (m), 
temperature, T ($^{\circ}$K), and kinetic energy, KE (m$^{2}$\,sec$^{-2}$) 
at 1000 mb.
  
\noindent {\bf Figure~11.} The same as in Figure 10 but at 700 mb.

\noindent {\bf Figure~12.}  Scattering diagram of daily coefficients (a0, a1) 
of linear regression  of HadAM3P field on Eta CCS model field of geopotential height 
at 1000 mb calculated over the all integration domain (top); time series of 
regression coefficients (a0, a1) (middle), time series of consistency index for 
these models (bottom).

\noindent {\bf Figure~13.} The same as in Figure 12 but for temperature at 1000 mb.

\noindent {\bf Figure~14.} Time spectra of mean (over the integration domain) 
geopotential height (top), temperature
(middle), and kinetic energy (bottom) at 1000 mb, provided by 
HadAM3P (solid) and Eta CCS model (dot-dashed) simulations.
\newpage
\begin{figure}[p]
\centerline{
      \includegraphics[angle=-90, width=85mm]{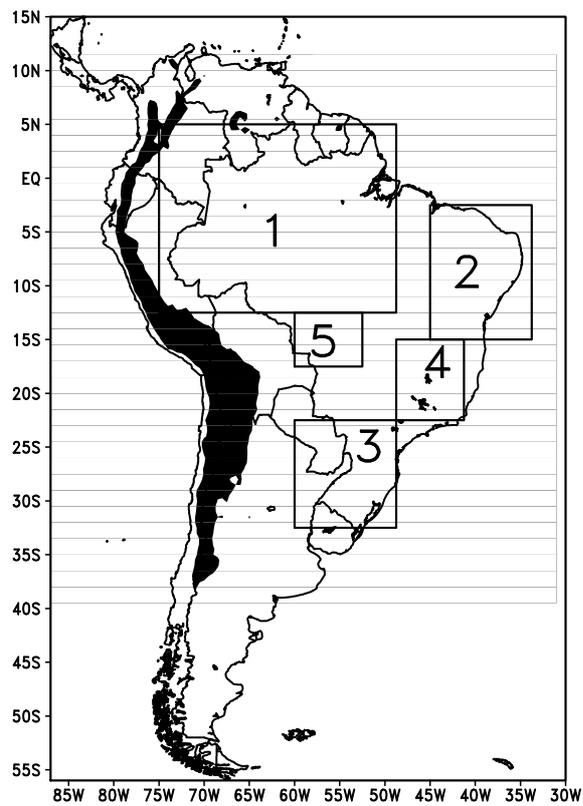}
      } 
      \caption[1]{\label{fig:fig1_Reg5}The regions over South America selected for the analysis:
Amazonia (1), Nordeste (2), Sul Brasil (3), Minas (4), 
Pantanal (5).
      }
\end{figure}
\clearpage
\begin{figure}[p]
\centerline{
      \includegraphics[angle=0, width=90mm]{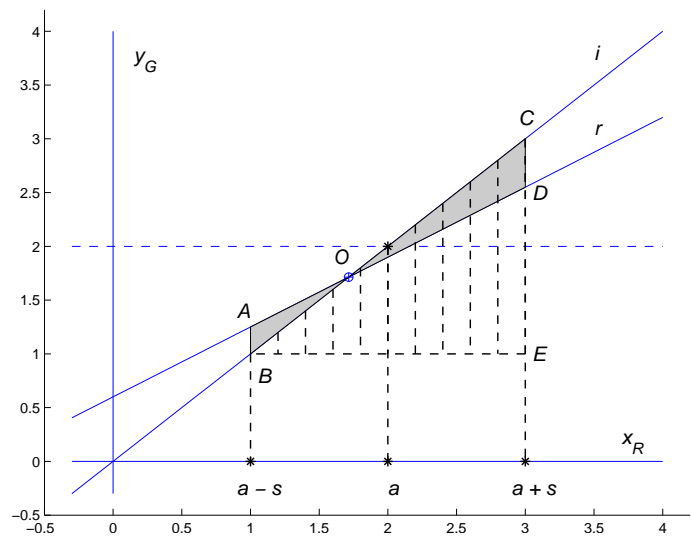}
      } 
      \caption[1]{\label{fig:fig2_risunok}Definition of consistency index by using
the coefficients of  linear regression of HadAM3P field on Eta CCS model field.
      }
\end{figure}
\clearpage
\begin{figure}[p]
\centerline{
      \includegraphics[angle=0, width=160mm]{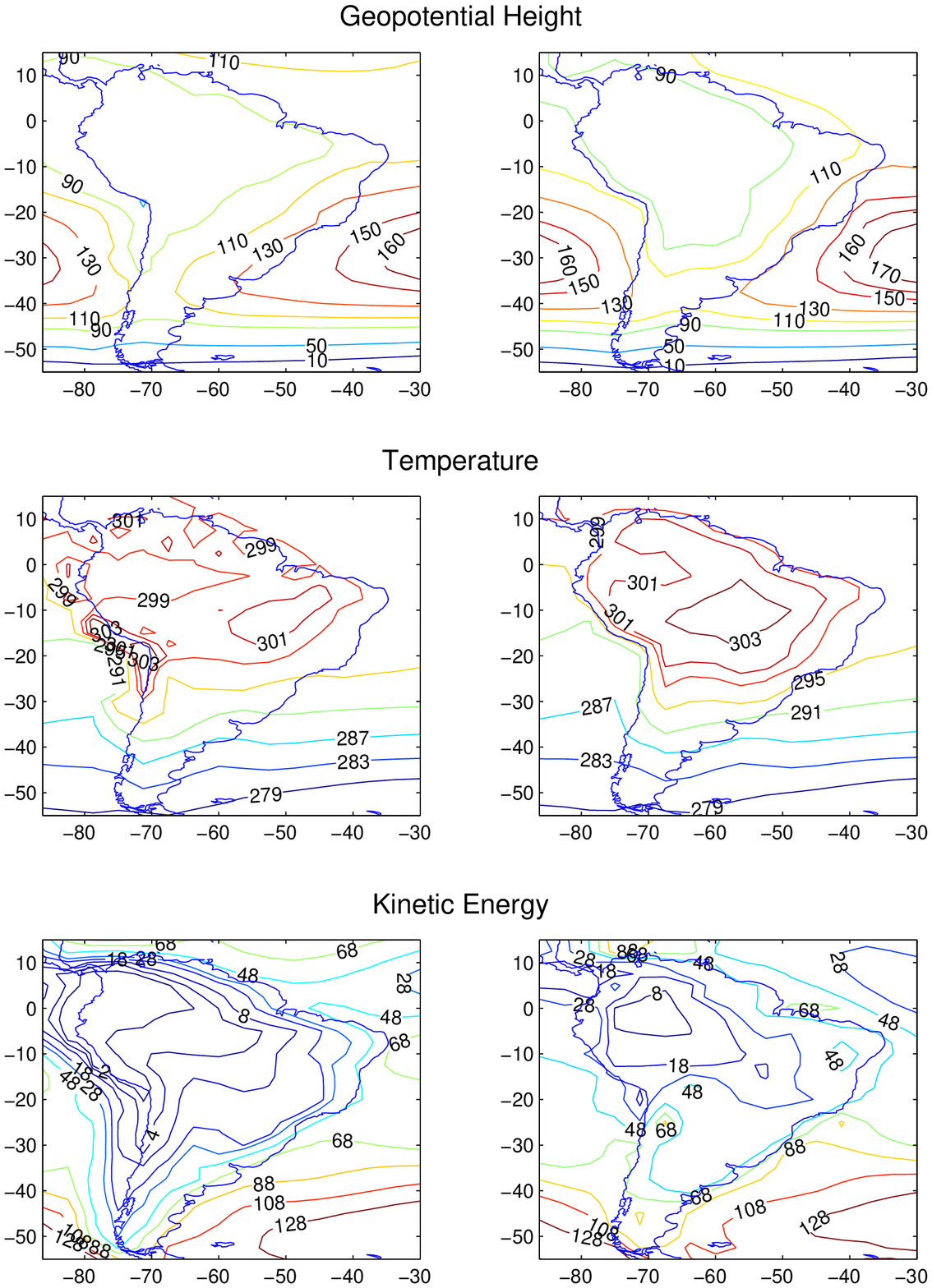}
      }
      \caption[1]{\label{fig:fig3_mean_fields_1000_0}Mean (1961-1990) fields 
of geopotential height(m), temperature ($^{\circ}$K), and kinetic 
energy (m$^{2}$\,sec$^{-2}$) at 1000 mb, provided by HadAM3P (left) 
and Eta CCS model (right) simulations.        
      }
\end{figure}
\clearpage
\begin{figure}[p]
\centerline{
      \includegraphics[angle=0, width=160mm]{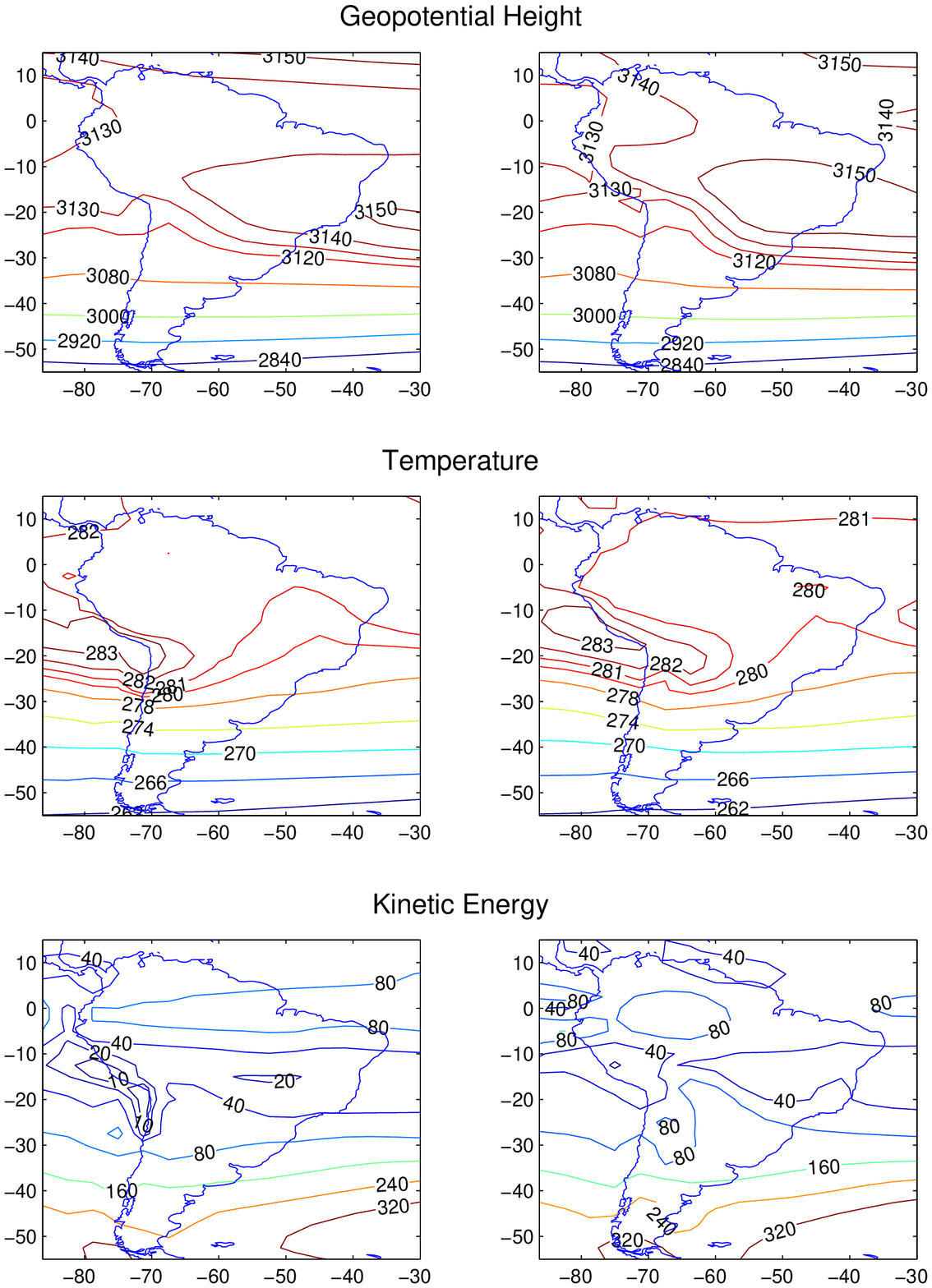}
      } 
      \caption[1]{\label{fig:fig4_mean_fields_700_0}The same as in Figure 2 but at 700 mb.
      }
\end{figure}
\clearpage
\begin{figure}[p]
\centerline{
      \includegraphics[angle=0, width=160mm]{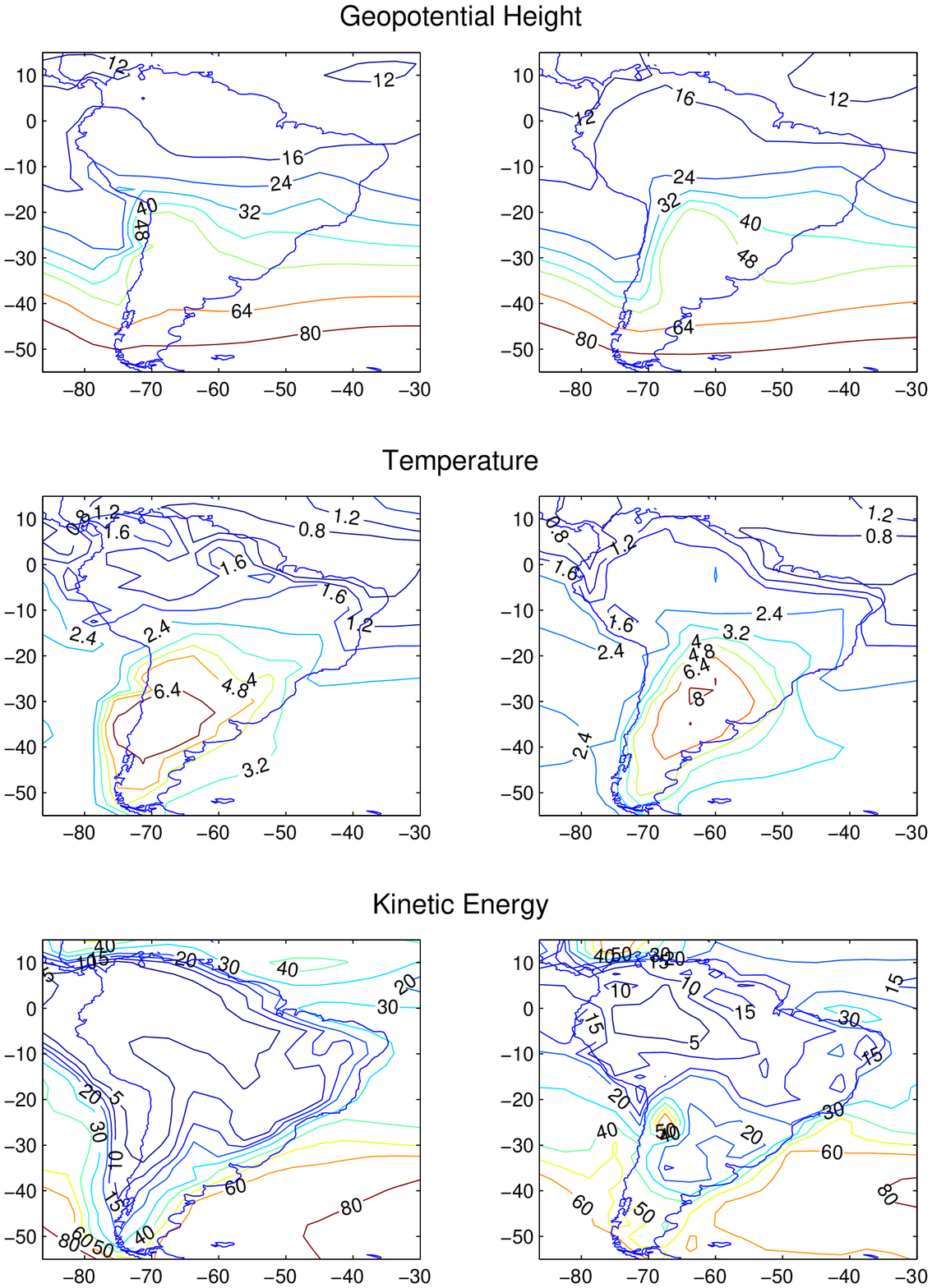}
      } 
      \caption[1]{\label{fig:fig5_stand_dev_1000_0} Mean (1961-1990) 
standard deviation fields of geopotential height (m), temperature ($^{\circ}$K), 
and kinetic energy (m$^{2}$\,sec$^{-2}$) at 1000 mb, provided by 
HadAM3P (left) and Eta CCS model (right) simulations.
      }
\end{figure}
\clearpage
\begin{figure}[p]
\centerline{
      \includegraphics[angle=0, width=160mm]{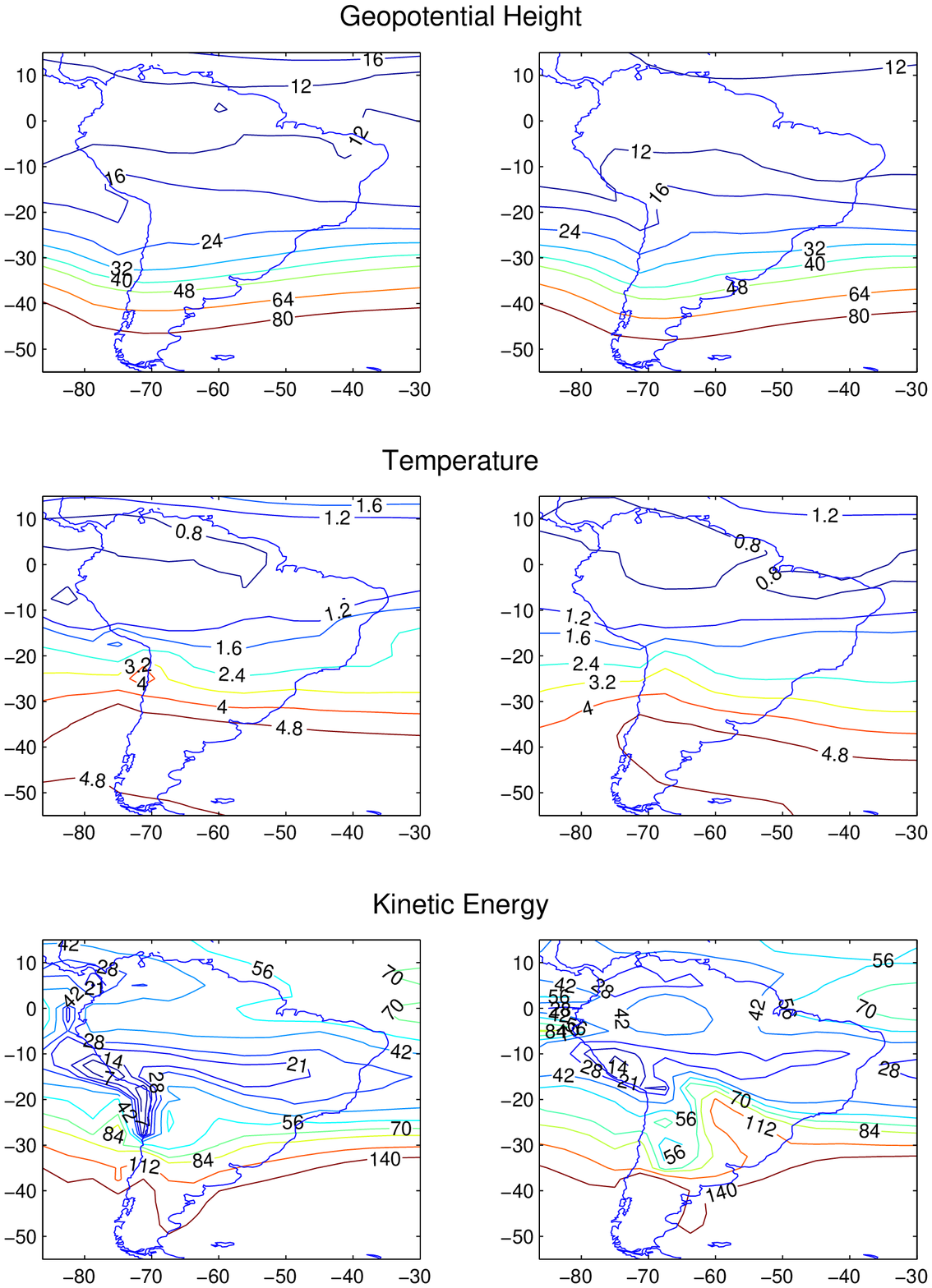}
      } 
      \caption[1]{\label{fig:fig6_stand_dev_700_0} The same as in Figure 4 but 
    at 700 mb.
      }
\end{figure}
\clearpage
\begin{figure}[p]
\centerline{
      \includegraphics[angle=0, width=150mm]{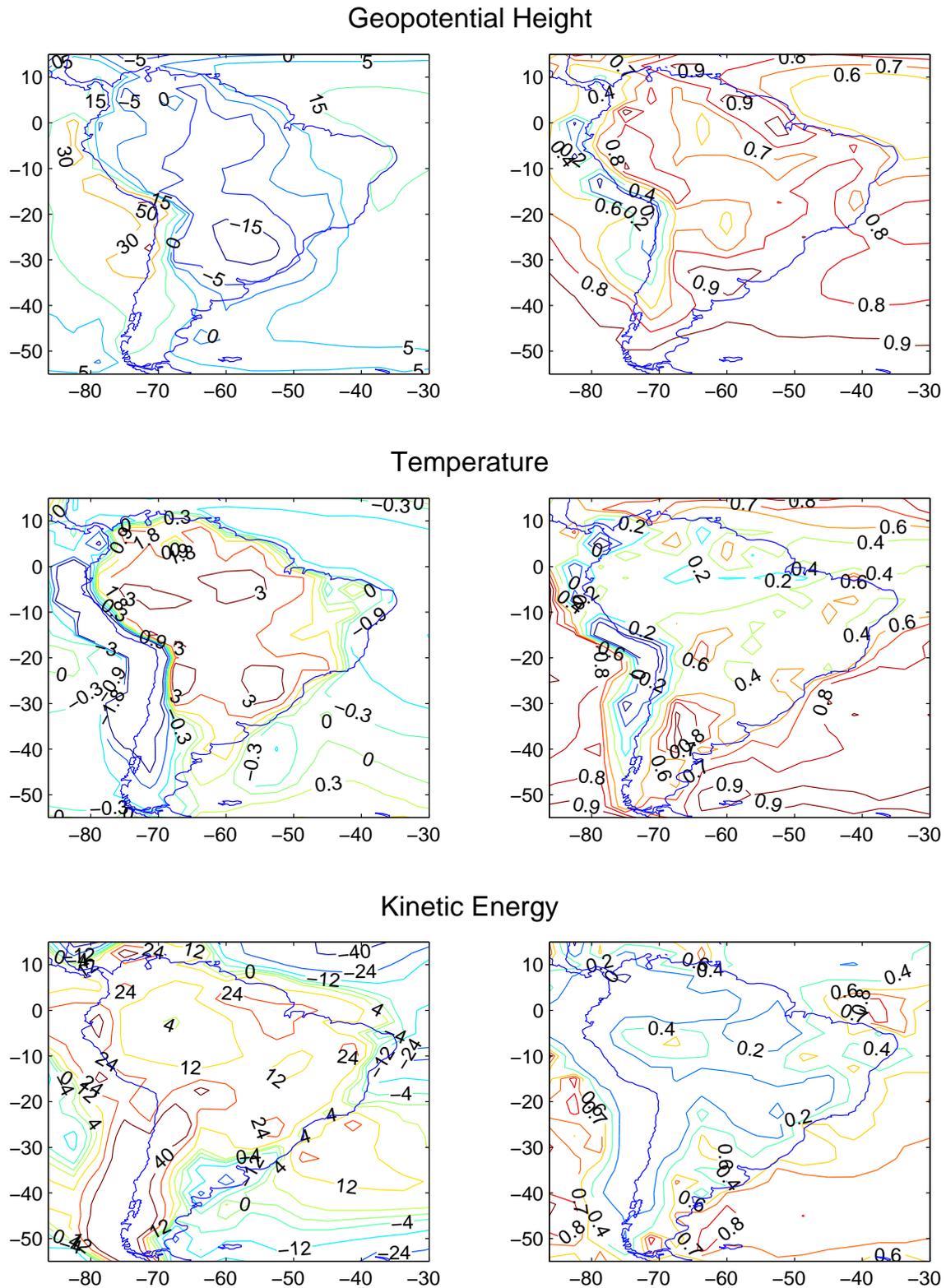}
      } 
      \caption[1]{\label{fig:fig7_mad_and_consist_ind_1000_0}Mean (1961-1990)
fields  of MAD (left),  calculated for HadAM3P and Eta CCS model fields 
of geopotential height (m), temperature ($^{\circ}$K), and  kinetic energy 
(m$^{2}$\,sec$^{-2}$) at 1000 mb, and consistency index between
HadAM3P and Eta CCS model(right), calculated for the same fields.
      }
\end{figure}
\clearpage
\begin{figure}[p]
\centerline{
      \includegraphics[angle=0, width=160mm]{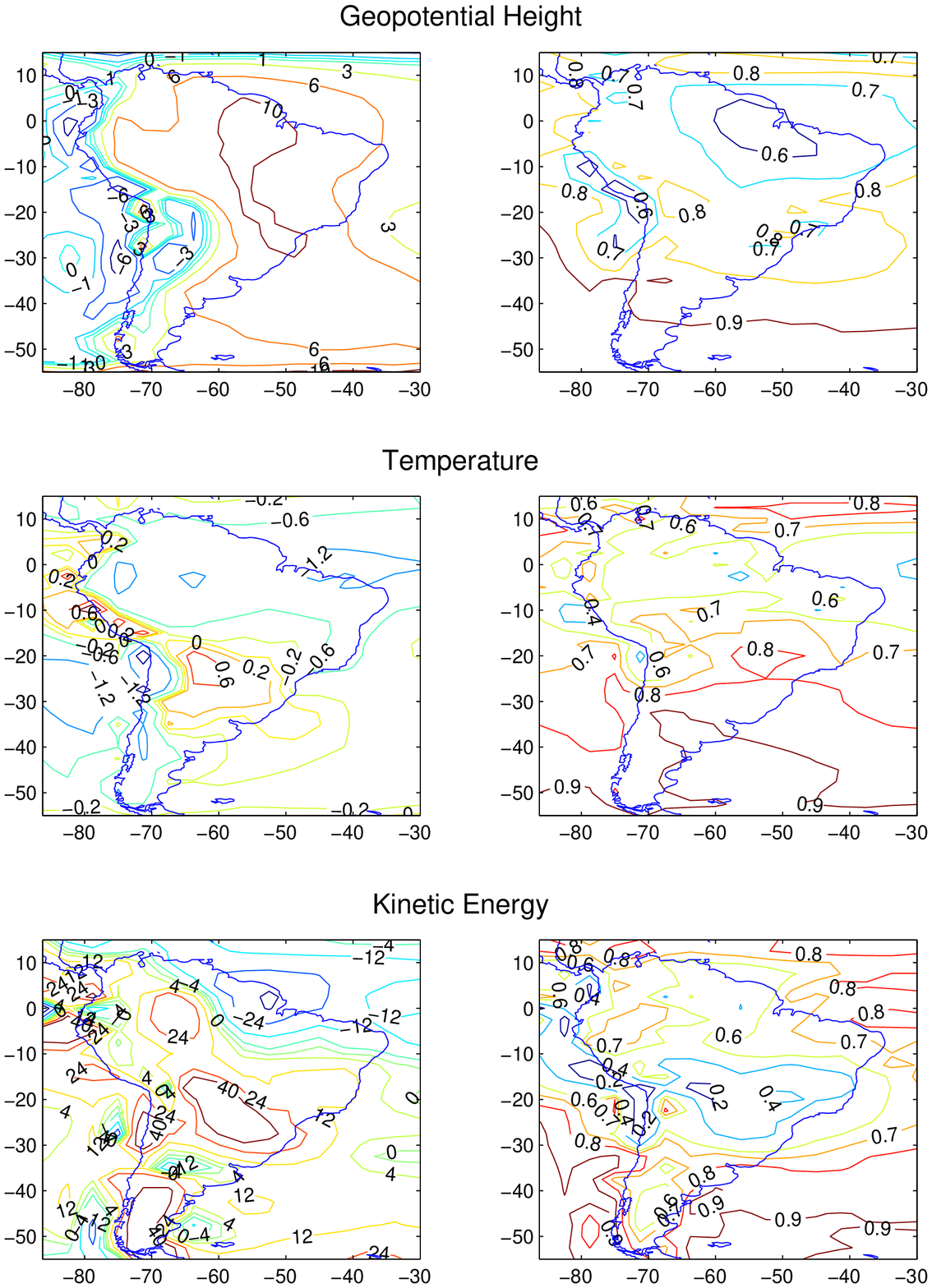}
      } 
      \caption[1]{\label{fig:fig8_mad_and_consist_ind_700_0} The same as in Figure 7 but 
      at 700 mb.
      }
\end{figure}
\clearpage
\begin{figure}[p]
\centerline{
      \includegraphics[angle=0, width=160mm]{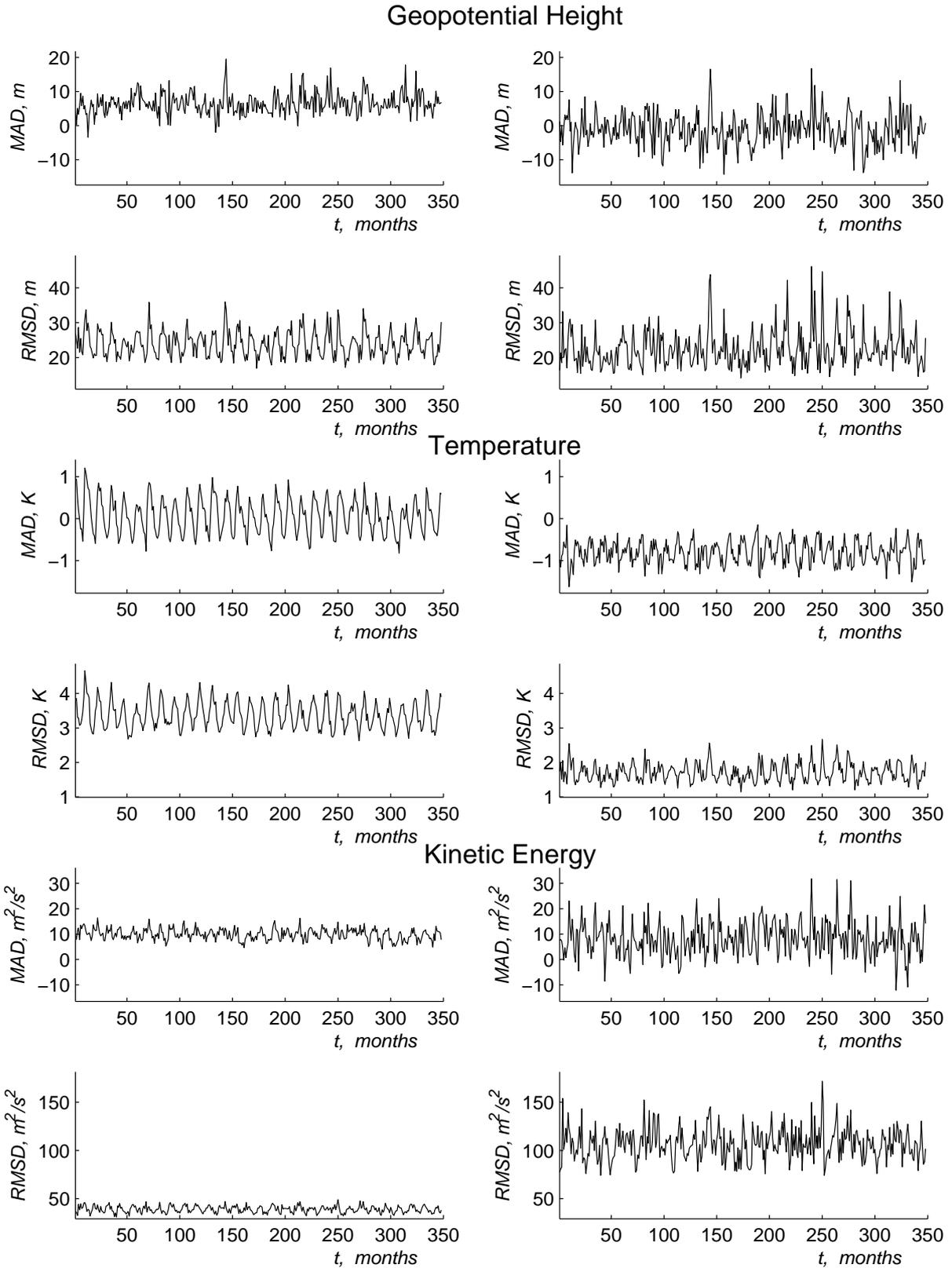}
      } 
    \caption[1]{\label{fig:fig9_mad_rmsd_1000_500_S_Am} Time series
of mean (over the integration domain) MAD and RMSD, 
calculated for HadAM3P and Eta CCS model fields of geopotential height (m), 
temperature ($^{\circ}$K), and kinetic energy (m$^{2}$\,sec$^{-2}$) 
at 1000 mb (left) and 500 mb (right).
      }
\end{figure}
\clearpage
\begin{figure}[p]
\centerline{
      \includegraphics[angle=0, width=110mm]{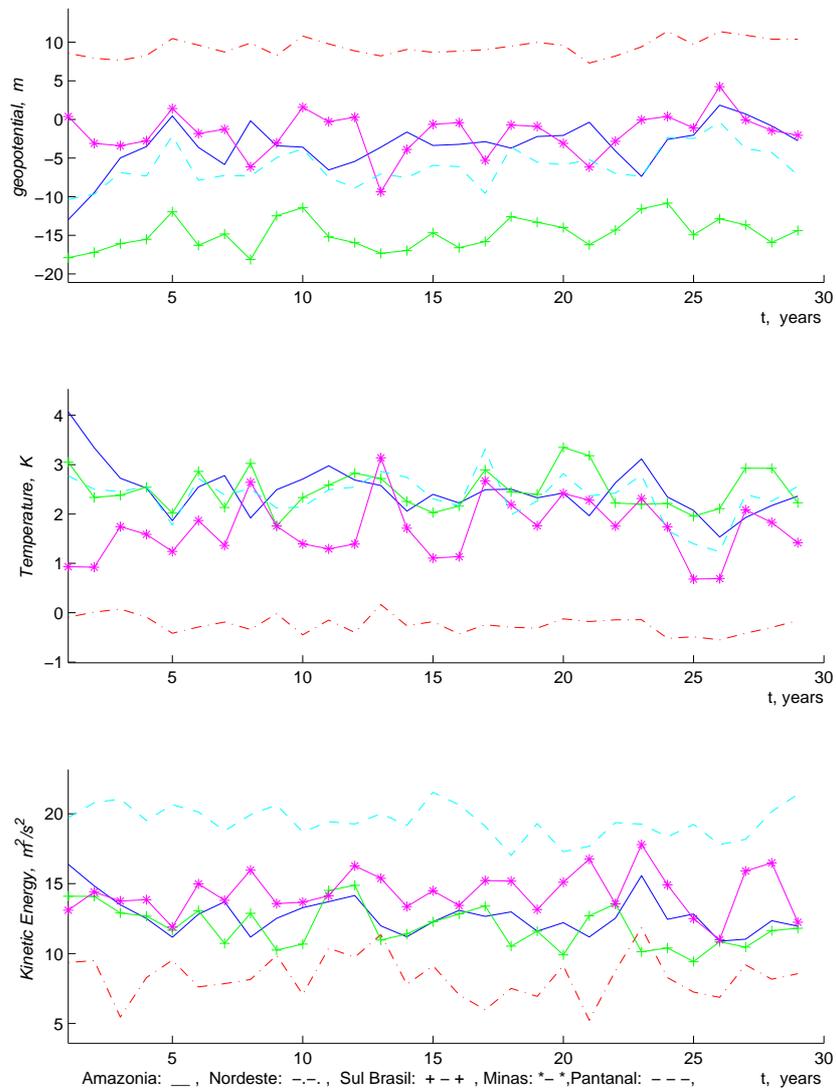}
      } 
      \caption[1]{\label{fig:fig10_mad_reg_1000}Time series of mean 
(over the regions shown in Figure 1) MAD, calculated for HadAM3P and 
Eta CCS model fields of geopotential height, G (m),  temperature, T 
($^{\circ}$K), and kinetic energy, KE (m$^{2}$\,sec$^{-2}$)  at 1000 mb.  
      }
\end{figure}

\clearpage
\begin{figure}[p]
\centerline{
      \includegraphics[angle=0, width=110mm]{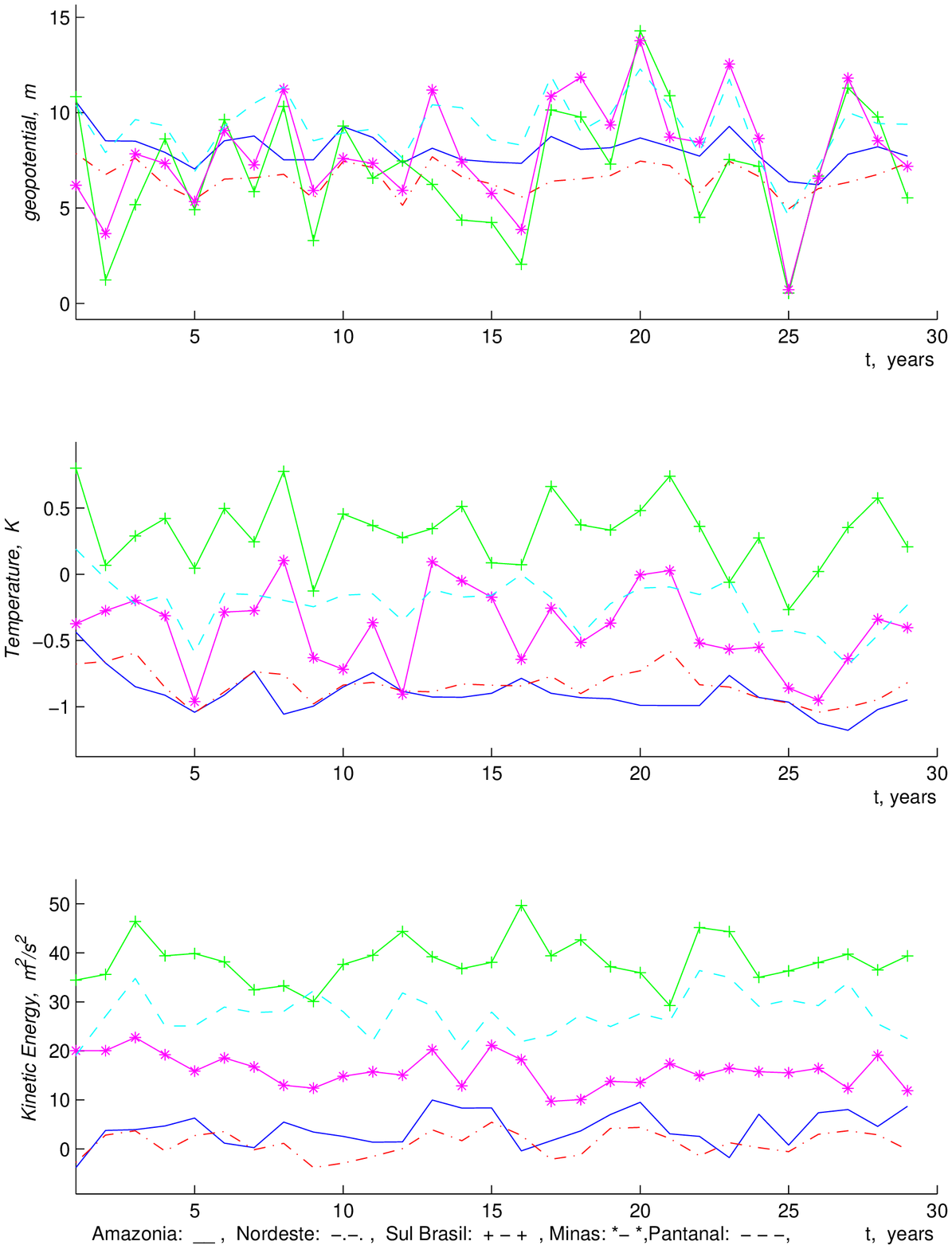}
      } 
      \caption[1]{\label{fig:fig11_mad_reg_700}The same as in Figure 10 but 
at 700 mb.
      }
\end{figure}

\clearpage
\begin{figure}[p]
\centerline{
      \includegraphics[angle=0, width=140mm]{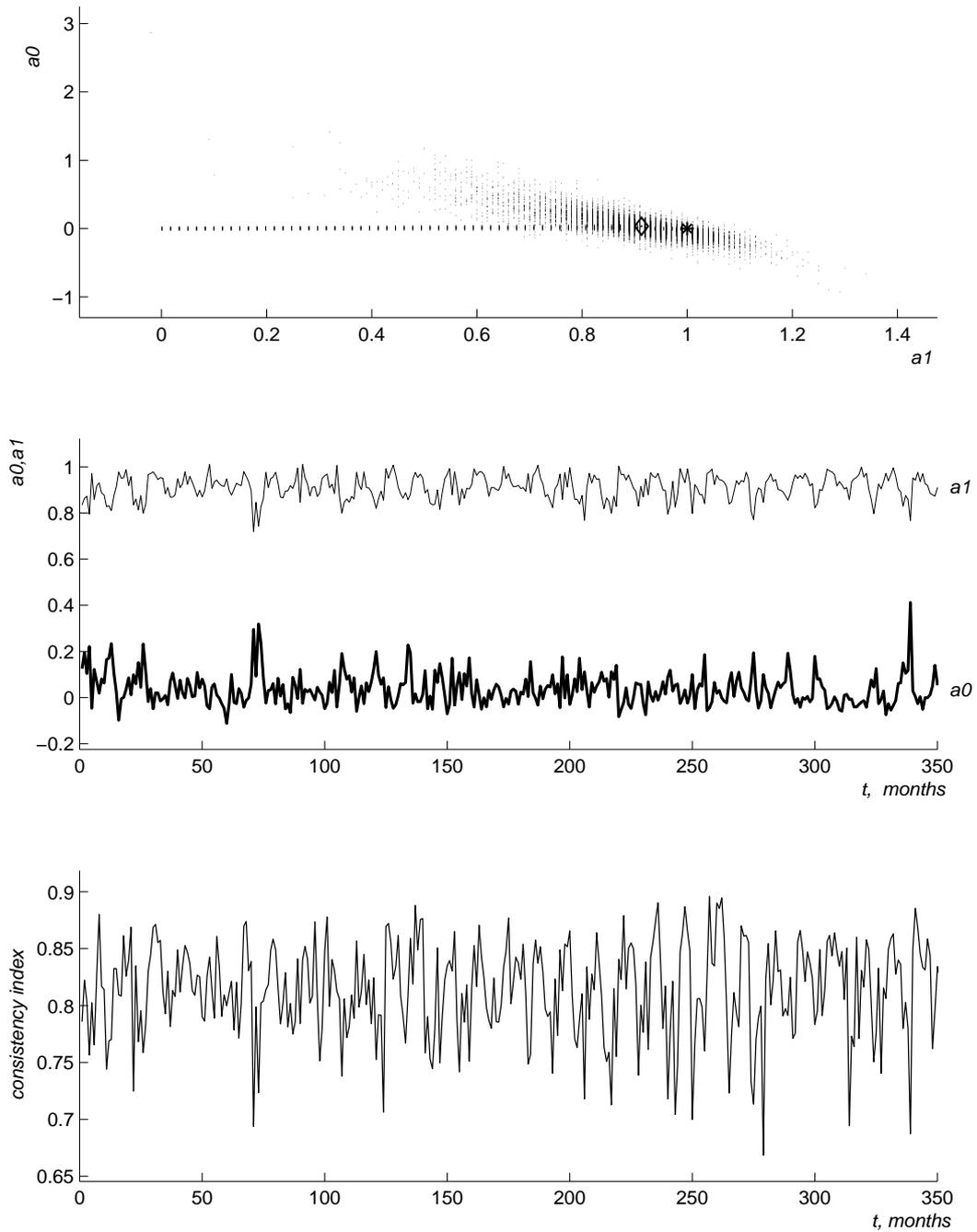}
      } 
      \caption[1]{\label{fig:fig12_coeff_lin_regr_h-sm_g1000} Scattering diagram of 
daily coefficients (a0, a1) 
of linear regression  of HadAM3P field on Eta CCS model field of geopotential height 
at 1000 mb calculated over the all integration domain (top); time series of 
regression coefficients (a0, a1) (middle), time series of consistency index for 
these models (bottom). 
      }
\end{figure}
\clearpage
\begin{figure}[p]
\centerline{
      \includegraphics[angle=0, width=140mm]{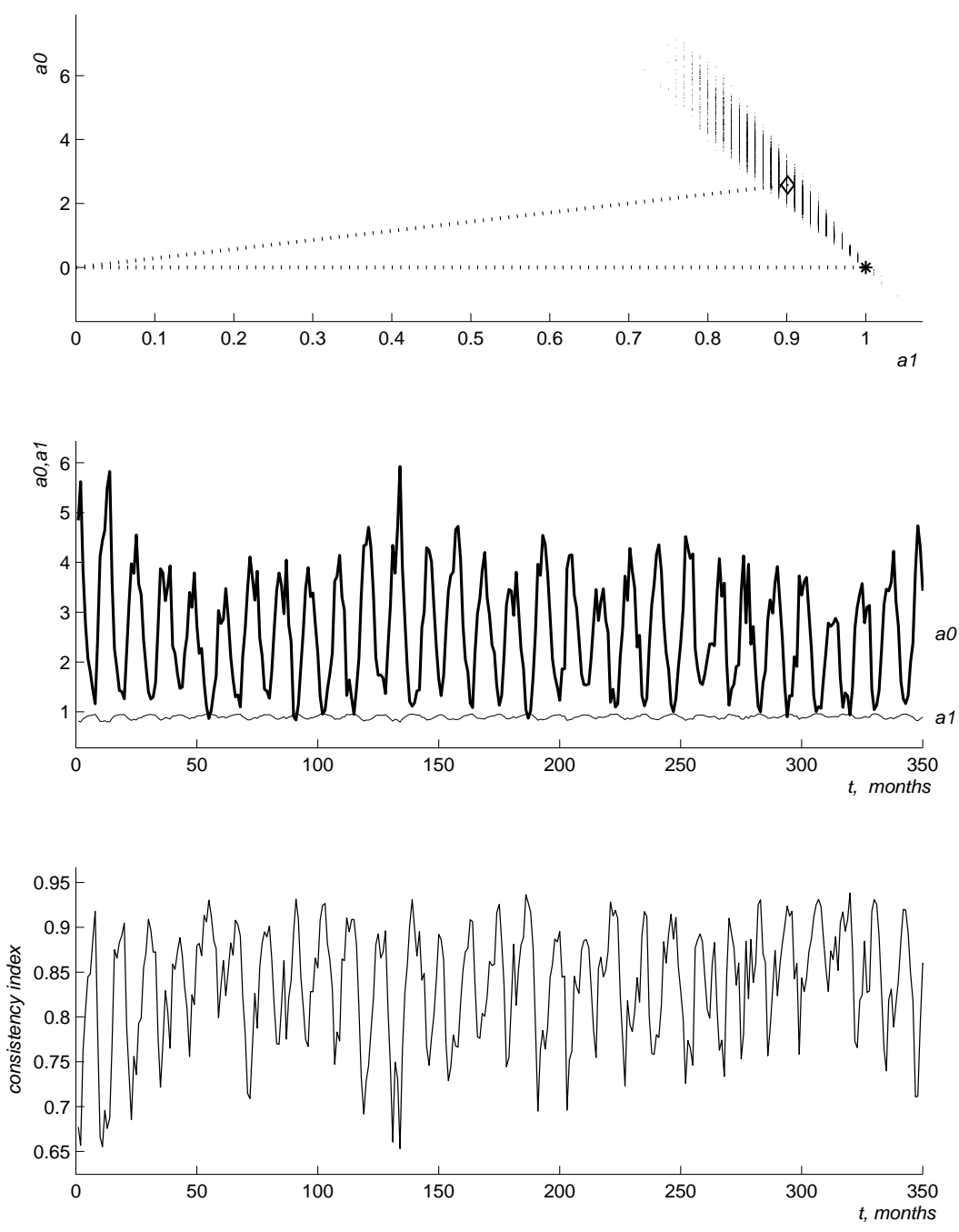}
      } 
      \caption[1]{\label{fig:fig13_coeff_lin_regr_h-sm_t1000} The same as 
in Figure 12 but for temperature at 1000 mb.  
      }
\end{figure}

\clearpage
\begin{figure}[p]
\centerline{
      \includegraphics[angle=0, width=160mm]{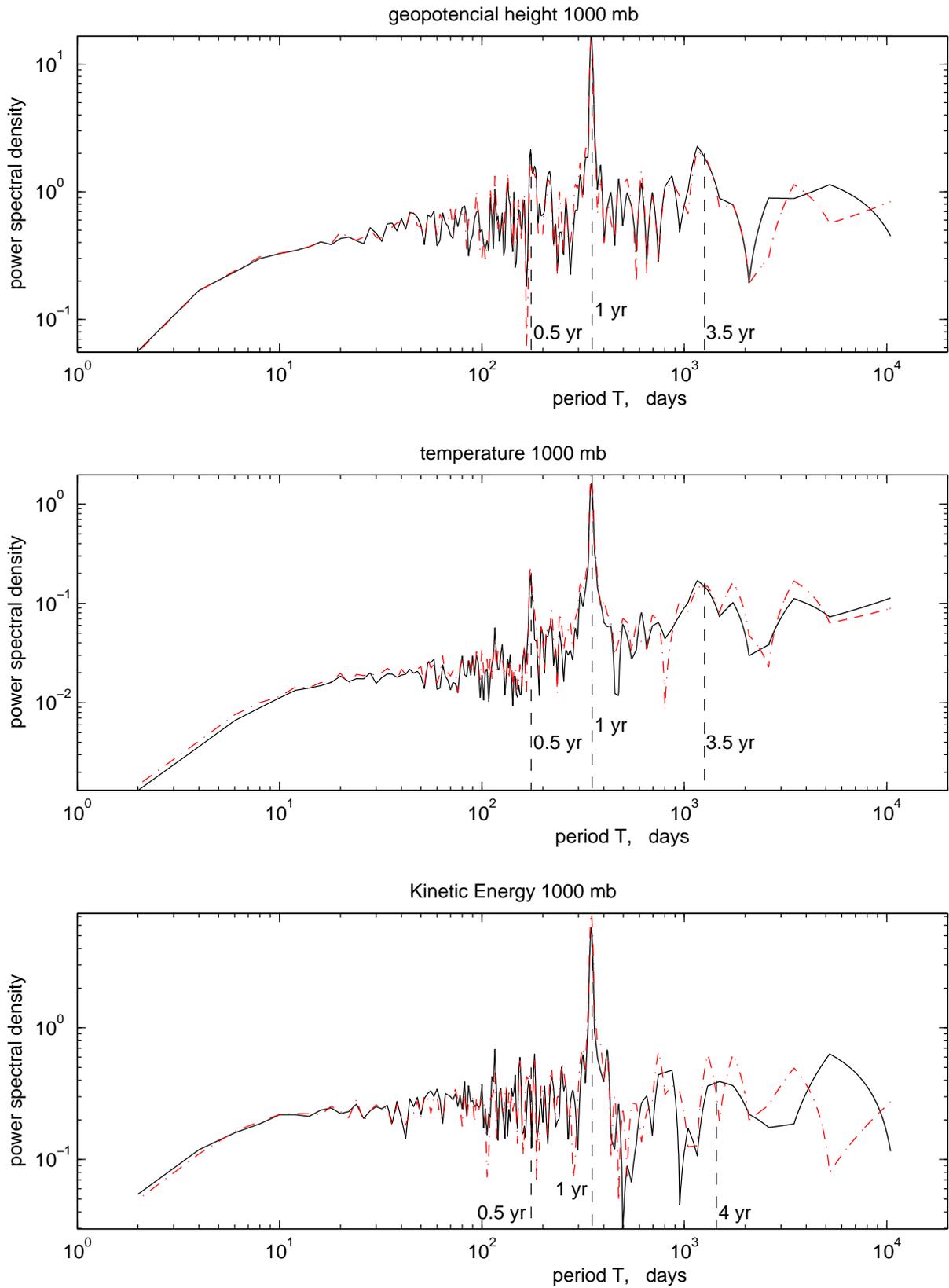}
      } 
      \caption[1]{\label{fig:fig14_spectr_america1000} Time spectra of mean 
(over the integration domain) geopotential height (top), 
temperature(middle), and kinetic energy (bottom) at 1000 mb, 
provided by HadAM3P (solid) and Eta CCS model (dot-dashed) simulations.
      }
\end{figure}
\clearpage
\noindent Table 1. Mean correlation coefficient (r), mean MAD , and mean RMSD
between the regional and global models time series of geopotential height (G),
temperature (T), and kinetic energy (KE) at 1000 mb and 500 mb, averaged over 
the integration domain (D) and over the 5 regions shown in Figure 1.    

\bigskip
\bigskip

\begin{tabular}{cccccccccc}
\hline
\hline 
%
 \multicolumn{1}{c}{} &
 \multicolumn{3}{c}{G}  &  
 \multicolumn{3}{c}{T}  &  
 \multicolumn{3}{c}{KE}    \\ \cline{2-10}

     Region    &
     r    &
     MAD    &
     RMSD   &
     r     &
     MAD    &
     RMSD    &	 
     r    &	 
     MAD    &	 
     RMSD      \\ [3pt] 
\hline	
  \multicolumn{10}{c}{Pressure level of 1000 mb} \\
  D   &  0.98  &  6   & 24   &  0.98  & 0.1   &  3.4   & 0.95   & 10   &  39  \\  	 
  1   &  0.95  &  -3  &  9   &  0.78  & 2.5   &  3.0   & 0.51   & 13   &  17  \\ 
  2   &  0.97  &  9   & 13   &  0.92  & -0.2  &  1.7   & 0.9    & 8    &  23  \\ 
  3   &  0.97  &  -15 & 25   &  0.96  & 2.5   &  4.2   & 0.83   & 12   &  27  \\ 
  4   &  0.95  &  -2  & 17   &  0.72  & 1.7   &  3.0   & 0.69   & 14   &  20  \\ 
  5   &  0.97  &  -6  & 14   &  0.64  & 2.4   &  3.5   & 0.79   & 20   &  22  \\ 
  
  \multicolumn{10}{c}{Pressure level of 500 mb}  \\ 
       
  D   &  0.97  &  -1  & 23   &  0.99  & -0.8  &  1.7   & 0.98   & 8    &  11  \\      
  1   &  0.97  &  -2  & 6    &  0.81  & -1.0  &  1.4   & 0.81   & 13   &  42  \\      
  2   &  0.94  &  -1  & 8    &  0.81  & -0.9  &  1.5   & 0.61   & 12   &  40  \\      
  3   &  0.89  &  3   & 26   &  0.97  & -1.0  &  1.8   & 0.93   &  7   &  111  \\      
  4   &  0.74  &  2   & 16   &  0.88  & -1.1  &  1.6   & 0.86   & 9    &  55  \\ 
  5   &  0.77  &  -1  & 10   &  0.79  &  -1.6 &  1.8   & 0.84   & 11   &  36  \\  [4pt]
     
\hline
\end{tabular}
\end{document}